\documentclass{aa}

\usepackage{graphicx}
\usepackage{amsfonts}
\usepackage{amsmath}
\usepackage{longtable}

\newcommand{\msun}{M_{\odot}}

\begin{document}

\title{Brown dwarfs in the Pleiades cluster : clues to the substellar
  mass function \thanks{Based on observations obtained at
    Canada-France-Hawaii Telescope}}

\author{E. Moraux, J. Bouvier, J.R. Stauffer \& J-C. Cuillandre}

\author{E. Moraux\inst{1}\and J. Bouvier\inst{1}\and 
  J.R. Stauffer\inst{2}\and J-C. Cuillandre\inst{3,4}}

\offprints{Estelle.Moraux@obs.ujf-grenoble.fr}


\institute{Laboratoire d'Astrophysique, Observatoire de Grenoble, B.P.
53, 38041 Grenoble Cedex 9, France
\and SIRTF Science Center, California Institute of Technology, Pasadena,
CA 91125, USA
\and Canada-France-Hawaii Telescope Corp., Kamuela, HI 96743, USA
\and Observatoire de Paris, 61 Av. de l'Observatoire, 75014 Paris, France}

\date{Received .../ Accepted ...}

\authorrunning{E. Moraux et al.}  
\titlerunning{Brown dwarfs in the Pleiades cluster}

\abstract{ We present the results of a 6.4 square degrees imaging
  survey of the Pleiades cluster in the $I$ and $Z$-bands. The survey
  extends up to 3 degrees from the cluster center and is $90\%$
  complete down to $I\simeq 22$. It covers a mass range from
  0.03$\msun$ to 0.48$\msun$ and yields 40 brown dwarf candidates
  (BDCs) of which 29 are new. The spatial distribution of BDCs is
  fitted by a King profile in order to estimate the cluster substellar
  core radius. The Pleiades mass function is then derived accross the
  stellar-substellar boundary and we find that, between $0.03\msun$
  and $0.48\msun$, it is well represented by a single power-law,
  $dN/dM\propto M^{-\alpha}$, with an index $\alpha=0.60\pm 0.11$.
  Over a larger mass domain, however, from 0.03$\msun$ to 10$\msun$,
  the mass function is better fitted by a log-normal function. We
  estimate that brown dwarfs represent about 25\% of the cluster
  population which nevertheless makes up less than 1.5\% of the
  cluster mass. The early dynamical evolution of the cluster appears
  to have had little effect on its present mass distribution at an age
  of 120~Myr. Comparison between the Pleiades mass function and the
  Galactic field mass function suggests that apparent differences may
  be mostly due to unresolved binary systems.
}

\maketitle


\section{Introduction}
%
The Initial Mass Function (IMF), i.e. the mass spectrum resulting
from complex physical processes at work during star formation, is a
formidable constraint for star formation models. Its determination at
low stellar and substellar masses is therefore one of the main
motivation for the rapidly expanding quest for brown dwarfs. Very low
mass stars and brown dwarfs are also prime candidates to investigate
the structure and dynamical evolution of large stellar systems, such
as e.g. clusters. Searches for the lowest mass, isolated objects have
been conducted in the galactic field, in young open clusters and in
star forming regions, brown dwarfs being brighter when younger. The
identification of substellar candidates is often based on color
magnitude diagrams (hereafter CMD) built from deep wide-field imaging
surveys. One of the major shortcomings of this selection method is the
contamination by older and more massive late-type field dwarfs which
may lie in the same region of the CMD.

The Pleiades cluster (RA=$3^{h}46.6^{m}$, Dec=$+24^{o}04'$) is an
ideal hunting ground for substellar objects. It is relatively nearby
with a distance of about 125 pc, with the apparent magnitude of
massive brown dwarfs bright enough for being easily detected on
intermediate-size telescopes. Its age of approximately 120 Myr
(Stauffer et al. 1998, Martin et al. 1998) makes the lithium
test particularly useful for identifying brown dwarfs~: at this age,
the lithium depletion boundary (in mass) coincides with the hydrogen
burning mass limit (hereafter HBML).  Also, the Pleiades is a rich
cluster with about 1200 members and its galactic latitude is
relatively high ($b=-23^{o}$) which minimizes the possible confusion
between members and red giants from the Galactic disk. Moreover, the
cluster motion ($\mu_{\alpha}=+19$ mas/yr, $\mu_{\delta}=-43$ mas/yr)
is large compared to that of field stars so that cluster kinematic
studies are a powerful way to recognize true members among the
photometric candidates.

\begin{table*}[htbp]
        \centering \leavevmode
\begin{tabular}{lccccc}
\hline
Survey & Area & Completeness & Nb of new & IMF index & Mass range \\
        & sq. degree & limit & BD candidates & & ($\msun$) \\
\hline
Stauffer et al. (1989) & 0.25 & $I\sim17.5$ & 4 & & 0.2-0.08 \\
Simons \& Becklin (1992) & 0.06 & $I\sim19.5$ & 22 &  & 0.15-0.045\\
Stauffer et al. (1994) & 0.4 & $I\sim17.5$ & 2 &  & 0.3-0.075\\
Williams et al. (1996) & 0.11 & $I\sim19$ & 1 &  & 0.25-0.045 \\
Festin (1997) & 0.05 & $I\sim21.6$ & 0 & $\le1$ &  0.15-0.035 \\
Cossburn et al. (1997) & 0.03 & $I\sim20$ & 1 &  & 0.15-0.04 \\
Zapatero et al. (1997) & 0.16 & $I\sim19.5$ & 9 & $1\pm0.5$ & 0.4-0.045 \\
Festin (1998) & 0.24 & $I\sim21.4$ & 4 & $\le1$ &  0.25-0.035 \\
Stauffer et al. (1998) & 1. & $I\sim18.5$ & 3 &  & 0.15-0.035\\
Bouvier et al. (1998)$^{a,b}$ & 2.5 & $I\sim22$ & 13 & $0.6\pm0.15$ & 0.4-0.04 \\
Zapatero-Osorio et al. (1999)$^b$ & 1. & $I\sim21$ & 41 &  & 0.08-0.035 \\
Hambly et al. (1999) & 36. & $I\sim18.3$ & 6 & $\le0.7$ &  0.6-0.06 \\
Pinfield et al. (2000)$^b$ & 6 & $I\sim19.6$ & 13 &  & 0.45-0.045 \\
Tej et al. (2002) & 7 & $K\sim15$ & & $0.5\pm0.2$ & 0.5-0.055 \\
Dobbie et al. (2002)$^{b,c}$ & 1.1 & $I\sim22$ & 10 & 0.8 & 0.6-0.030 \\
\hline
\end{tabular}

\parbox{0.9\textwidth}{%
  $^a$BD candidates have been confirmed by Mart\'{\i}n et al. (2000)
  and Moraux et al. (2001) and the IMF index has been revised to
  $0.51\pm0.15$.

$^b$Jameson et al. (2002) have compiled these surveys and using new
infrared data they find that the Pleiades mass function is well
represented by a power law with index $\alpha=0.41\pm0.08$ for
$0.3\msun\ge M\ge 0.035\msun$.

$^c$These authors used the results from Hodgkin \& Jameson (2000) and Hambly 
et al. (1999) to conclude that the $\alpha=0.8$ power law is appropriate from 
$0.6\msun$ down to $0.03\msun$.\medskip

}
        \caption{Previous imaging surveys of the Pleiades conducted
          for searching brown dwarfs.}
        \label{bdsurveys}
\end{table*}

To date, several surveys have been conducted in this cluster. They are
summarized in Table~\ref{bdsurveys} in term of covered area,
completeness limit and number of new brown dwarfs candidates. Only
surveys with a magnitude limit larger or equal to $I\simeq17.5$ are
listed here (the HBML corresponds to $Ic=17.8$, Stauffer et al. 1998).
Starting in 1997\footnote{Prior to 1997, mass function estimates were
  based on strongly contaminated samples of brown dwarf candidates.},
several groups reported estimates for the Pleiades mass function in
the upper part of the substellar domain, often represented by a single
power law $dN/dM \propto M^{-\alpha}$ with $0.5\leq\alpha\leq 1.0$
(see Table~\ref{bdsurveys}).


While these various estimates agree within uncertaintites, the
substellar samples on which they rely are still relatively small and,
in some cases, not fully corrected for contamination by field dwarfs.
Some surveys are also quite limited spatially, and the derived mass
function may not be representative of the whole cluster. In order to
put the determination of the Pleiades substellar mass function on
firmer grounds, we performed a deep and large survey of the Pleiades
cluster in the $I$ and $Z$-band, using the CFH12K camera at the
Canada-France-Hawaii telescope. The survey covers 6.4 square degrees
and reaches up to 3 degrees from the cluster center. It encompasses a
magnitude range $I\simeq 13.5-24$, which corresponds to masses from
0.025$\msun$ to 0.45$\msun$ at the distance and age of the Pleiades.

In section 2, we describe the observations and the data reduction.
Results in the stellar and substellar domains are presented in Section
3. Forty brown dwarf candidates are identified, 29 of which are new
discoveries. We discuss contamination of the photometric samples by
field stars in the stellar and substellar domains, investigate the
radial distribution of substellar objects, and derive the cluster
substellar mass function. In Section 4, we discuss the overall shape
of the cluster mass function over the stellar and substellar domains,
the possible consequences of early cluster dynamical evolution on its
present mass function and the implications for the brown dwarf
formation process. A comparison is made between the Pleiades and the
Galactic field mass functions, which reveals significant differences
at the low mass end, a large part of which is probably attributable to
unresolved cluster binaries.

\section{Observations and data reduction}
\begin{figure}[htbp]
  \centering 
  \leavevmode 
  \includegraphics[width=0.9\hsize]{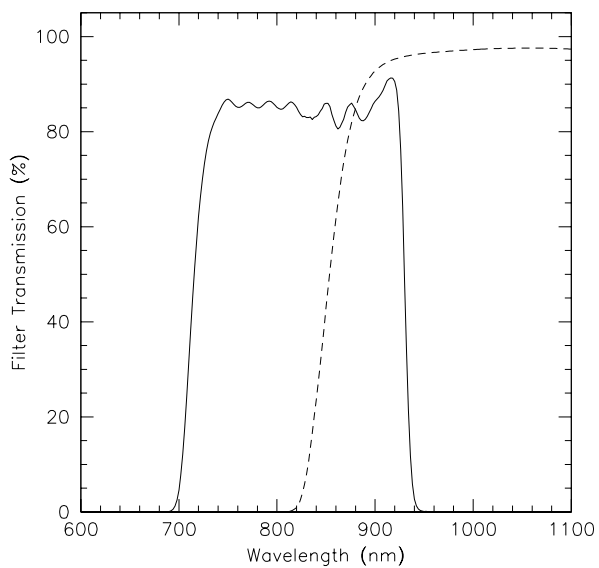}
  \caption{Filter profiles. The solid line corresponds to the $I$ Mould
    filter and the dashed line to the $Z$ filter. The cut-off in the
    $Z$-band is imposed by the detector pass band.}
  \label{filter}
\end{figure}

\begin{figure}[htbp]
  \begin{center}
    \leavevmode \includegraphics[width=0.9\hsize]{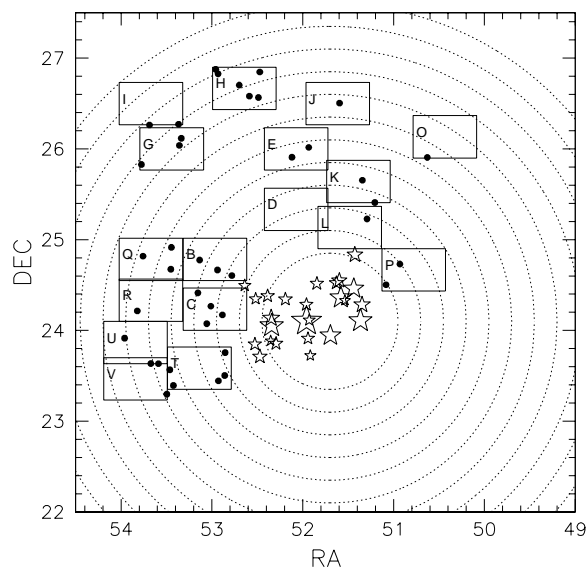}
  \end{center}
    \caption{Location of the 17 fields relative to the bright
      Pleiades members shown as stars. The size of the symbols depends
      on the brightness of the sources. The units are degrees on both
      axes.  Overplotted are the circles of radii from 0.75 to 3.5
      degrees centered on the cluster center. The dots represent the
      brown dwarf candidates detected in our survey. The total area
      covered by the survey amounts to 6.4 square degrees and reaches
      up to 3 degrees from the cluster center.}
    \label{fieldlocation}
\end{figure}

The $I$ and $Z$ observations were made using the CFH12K mosaic camera
(Cuillandre et al. 2001) at the prime focus of the
Canada-France-Hawaii telescope, from December 17 to 19, 2000. The
camera is equipped with an array of twelve $2048\times4096$ pixel CCDs
with $15{\mu}m$ size pixels, yielding a scale of 0.206''/pixel and a
field of view of 42'$\times$28'. The $I$ Mould and $Z$ filter
  profiles are shown in Fig~\ref{filter}. We obtained images of 17
fields in the Pleiades whose coordinates are given in
Table~\ref{fieldcoor} and the location on the sky relative to the
brightest Pleiades members shown in Fig~\ref{fieldlocation}. Fields
were selected in order to avoid the brightest stars which would
produce reflection halos and scattered light on the detectors, i.e., a
higher and non uniform background which is difficult to remove. We
also avoided the southwestern region of the cluster where a small CO
cloud causes relatively large extinction. The observing conditions
were photometric with a $I$-band seeing of typically 0.5'' FWHM, but
occasionally up to 1.1'' FWHM.

\begin{table}[htbp]
    \centering \leavevmode 
\begin{tabular}{llcccc} 
\hline 
Field & RA(2000) & DEC(2000) \\
 & (h m s) & ($^{o}$ ' '') \\
\hline
Pl-B & 03:51:55 & 24:47:00 \\
Pl-C & 03:51:55 & 24:14:00 \\
Pl-D & 03:48:00 & 25:20:00 \\
Pl-E & 03:48:00 & 26:00:00 \\
Pl-G & 03:54:00 & 26:00:00 \\
Pl-H & 03:50:30 & 26:40:00 \\
Pl-I & 03:55:00 & 26:30:00 \\
Pl-J & 03:46:00 & 26:30:00 \\
Pl-K & 03:45:00 & 25:38 30 \\
Pl-L & 03:45:25 & 25:08:00 \\
Pl-O & 03:40:50 & 26:08:00 \\
Pl-P & 03:42:20 & 24:40:00 \\
Pl-Q & 03:55:00 & 24:47:00 \\
Pl-R & 03:55:00 & 24:20:00 \\
Pl-T & 03:52:40 & 23:25:00 \\
Pl-U & 03:55:45 & 23:52:00 \\
Pl-V & 03:55:45 & 23:28:00 \\
\hline 
\end{tabular}
    \caption{Coordinates of the 17 different pointings observed with
      the CFH12K camera.}
    \label{fieldcoor}
\end{table}

\begin{figure}[htbp]
        \centering
        \leavevmode
          \includegraphics[width=0.9\hsize]{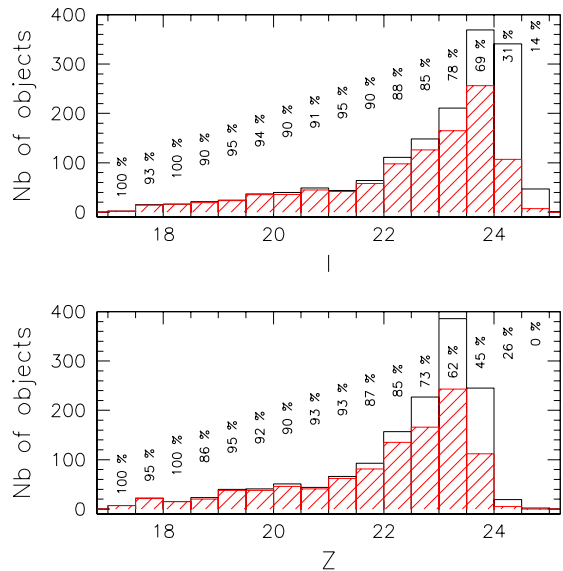}
        \caption{The plain histograms represent the number of sources
          per magnitude bin $I$ (top) or $Z$ (bottom) detected in the
          field Pl-U and located in the region which overlaps with
          field Pl-V. Hatched histograms correspond to the number of
          objects detected in both fields. The percentage of matched
          sources is given and indicates that the survey is about 90\%
          complete down to $I\simeq Z\simeq 22$.}
        \label{completeness}
\end{figure}

Short and long exposures were taken for each field in order to cover a
large magnitude range. We obtained a set of four images in both $I$
and $Z$ filters~: one 10 sec exposure plus three 300 sec exposures in
the $I$-band and one of 10 sec exposure plus three of 360 sec
exposures in the $Z$-band.  The short exposures encompass the
magnitude range $13.5 \le I \le 21.5$ whereas the long ones cover $17
\le I \le 24$. A comparison between the number of stellar like objects
detected in long exposures of overlapping fields indicates that the
survey is about 90\% complete down to $I\simeq22$ (see
Fig~\ref{completeness}). Note that the completeness limit of the short
exposures ($I \simeq 19.5$) is much larger than the saturation limit
of the long exposures ($I \simeq 17$), so that the survey completely
samples the continuous mass range from $0.03\msun$ to $0.45\msun$ in
the Pleiades cluster.

\subsection{Photometry}

Each mosaic consists of 12 CCD images which were reduced and analysed
separately. The images were overscan corrected, debiased and
flat-fielded. The flats were normalized to a reference CCD to retain
the appropriate relative scaling between chips.  The same photometric
zero-point can thus be used for all the CCDs of a mosaic.  The images
were also fringe corrected using patterns derived from a smoothed
combination of more than 20 images in each band.  Then, images of
similar exposure time for a given field were stacked together using an
optimal CCD clipping to remove the artefacts, and then taking the
average of the remaining final stack for each pixel to preserve the
photometry.

To detect all the sources in the frames, we averaged all long exposure
$I$ and $Z$ images of a field previously shifted to the same location
and used the automatic object-finding algorithm from the Sextractor
package (Bertin \& Arnouts 1996). Then, a PSF-fitting photometry on
the $I$ and $Z$ images was performed for all the detected objects
using the PSFex package.  We discriminated between point-like and
extended or corrupted sources using the FWHM distribution of all the
detected objects on each field. In practice, we defined conservative
lower and upper limits around the well defined stellar peak of this
distribution and rejected all the sources located outside these
boundaries.

\begin{figure}[htbp]
        \centering
        \leavevmode
          \includegraphics[width=0.9\hsize]{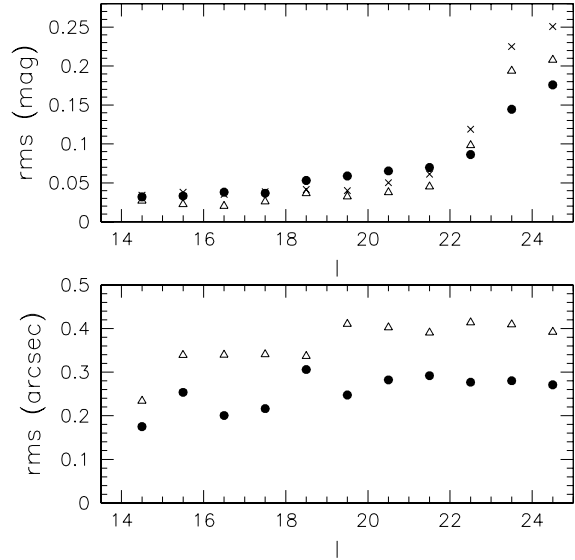}
          \caption{{\it Top~:} Photometric error rms plotted as a
            function of $I$ magnitude. The dots correspond to the
            error on the magnitude $I$, the open triangles to the
            error on $Z$ and the crosses to the error on $I-Z$; {\it
              Bottom~:} Astrometric error rms shown as a function of
            $I$ magnitude. The dots correspond to the error on the
            right ascension and the triangles denote the declination
            error.}
        \label{rms}
\end{figure}

Photometric standard Landolt fields SA98 and SA113 were observed and
reduced in the same way as the science images. Three successive
exposures of SA98 were obtained with an incremental offset of several
arcminutes in RA. Thus, common sets of photometric standard stars were
observed on every CCD of the mosaic and we checked that the
photometric zero-point in $I$-band was the same for each CCD~: we did
find a scatter of only 0.03 mag. The photometry in the $Z$-band is
sensitive to detector pass band and Landolt does not give $Z$
magnitudes for his standards. Thus we selected unreddened A0 standards
and set their $Z$ magnitude equal to their $I$ magnitude assuming that
their $I-Z$ color was zero (by definition of an A0 star). Thanks to
the several exposures of SA98 shifted of a few arcminutes, those A0
stars were observed on different chips so that we could derive $Z$
zero-points for some of the CCDs. We found a scatter of only 0.04 mag
and we then used the mean value as a global zero-point, assuming that
it could be used for all the CCDs as in the $I$-band. We verified this
assumption \textit{a posteriori} by measuring the difference between
the magnitudes of the same objects detected in the common region of 2
overlapping pointings. We thus verified that there was no systematic
error and estimated the photometric rms error up to the completeness
limit ($I\simeq 22$, see Fig.~\ref{rms}, top panel) which amounts to
0.07 mag or less at $I\le 22$.

\subsection{Astrometry}

In order to obtain accurate coordinates for cluster brown dwarf
candidates, we had to derive an astrometric solution for each CCD of
the mosaic.
We used the CFHT's Elixir package (Magnier \& Cuillandre 2002) to
compute the astrometric solution for each image.  The algorithm
calculates the celestial coordinates for all the detected objects
using the approximate solution given by the header, compares them with
the USNO2 catalog and refines the solution.  Most of the USNO2 stars
were saturated in our long exposure images so that we had to use lists
of refined coordinates derived from the short exposures as a reference
catalog.

We estimated the astrometric error by comparing coordinates of stars
present in overlapping regions and found an accuracy better than 0.5
arcsecond. The astrometric rms error is shown in Fig~\ref{rms} (bottom
panel) as a function of $I$ magnitude.

\section{Results}

The $(I,I-Z)$ color magnitude diagrams of the point-like objects
contained in the 17 Pleiades fields are shown in Fig~\ref{cmdshort}
and \ref{cmdlong}.  The short and long exposures have been analysed
separately corresponding respectively to the stellar and the
substellar domains.  In both cases we present our photometric
selection of candidates before dealing with the field star
contamination.  We examine the spatial distribution of cluster members
and attempt to measure their core radius. We then use these estimates
to derive the Pleiades mass function.

\subsection{Stellar domain}

\begin{figure}[htbp]
  \centering \leavevmode
  \includegraphics[width=0.9\hsize]{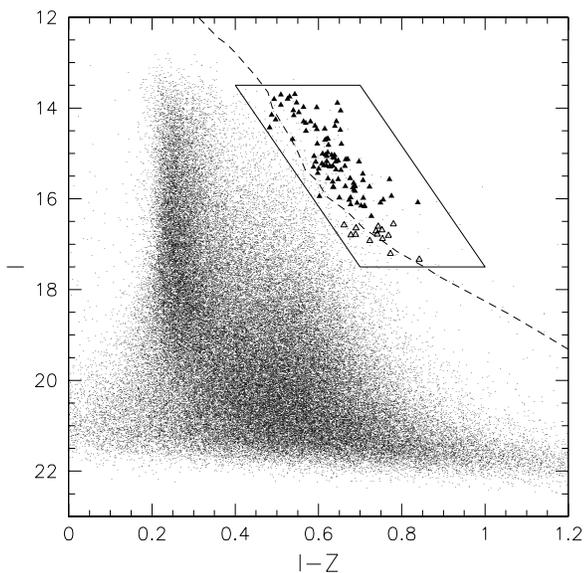}
    \caption{($I,I-Z$) color-magnitude diagram for the short exposures.
      The dashed line is the 120 Myr isochrone from the NEXTGEN models
      of Baraffe et al. (1998) shifted to the Pleiades distance. The
      region corresponding to our photometric selection for stellar
      candidates corresponds to the box. Objects recovered by 2MASS
      and having a membership probability $p$ based on their proper
      motion larger than 0.1 (Adams et al. 2001) are shown as filled
      triangles. Candidates too faint to be found in 2MASS but
      recovered by Hambly et al. (1999) and having a proper motion
      within $1\sigma$ of the cluster motion are indicated as open
      triangles.  }
    \label{cmdshort}
\end{figure}

The $(I,I-Z)$ color magnitude diagram for the short exposure images of
our survey is presented in Fig~\ref{cmdshort}. The 120 Myr isochrone
from the NEXTGEN models of Baraffe et al. (1998) shifted to the
Pleiades distance ($(m-M)_{o}= 5.53$) is shown as a dashed line.  On
the basis of the location of this theoretical isochrone, we made a
rather conservative photometric selection to include all possible
stellar members between $I=13.5$ and $I=17.5$. This selection
corresponds to the box drawn in Fig~\ref{cmdshort}.

To remove all the contaminating field stars from our sample, we
compared our list of candidates to the results of other large Pleiades
surveys.  Adams et al. (2001) performed a large search for Pleiades
stellar members using the photometry from the Two Micron All Sky
Survey (2MASS) and proper motions determined from Palomar Observatory
Sky Survey (POSS) plates. This search extends to a radius of 10$\degr$
around the cluster center, well beyond the tidal radius, which means
that it covers the complete cluster area. The completeness limit of
the POSS plates is $I\sim16.5$, i.e.  $0.1\msun$. The authors analysed
the proper motion of all the objects previously selected on the basis
of their 2MASS JHK photometry and defined a membership probability $p$
(see Adams et al.  2001 for details).  We cross-correlated our list of
stellar candidates with the list of all the sources analysed by Adams
et al. (2001) and we kept all the objects with $p>0.1$ so as to
minimize the non-member contamination down to $I=16.5$ (Adams' survey
completeness limit). All those sources are shown as filled triangles
in Fig~\ref{cmdshort}.

For stars fainter than $I>16.5$ we compared our results with those
from Hambly et al. (1999). They used photographics plates from the
United Kingdom Schmidt Telescope to construct a $6\degr \times 6\degr$
proper motion survey centered on the Pleiades.  To minimize the
contamination, we chose all the objects out of our photometric
candidates having proper motion within $1\sigma$ ($\simeq 20$ mas/yr)
of the known cluster motion ($\mu_{\alpha}=+19$ mas/yr,
$\mu_{\delta}=-43$ mas/yr, Robichon et al. 1999). They are indicated
as open triangles in Fig~\ref{cmdshort}.

We stopped our stellar selection at $I=17.5$ corresponding about to
Hambly's survey completeness limit but also to the HBML.  A short list
of those very probable low mass stellar members in our survey is
presented in Table~\ref{starcoo}.  We considered that the residual
contamination of this sample is low enough to be neglected. The
analysis of the fainter objects, i.e. substellar candidates, has been
done from the long exposure images and is explained hereafter.

\begin{table}[htbp]
        \centering \leavevmode
\begin{tabular}{lllll}
\hline
No & $I$ & $I-Z$ & $RA_{J2000}$ & $DEC_{J2000}$ \\
    &     &       & (h m s)    & (${}^{o}$ ' '')  \\
\hline 
1 & 13.69 & 0.54 & 3:51:11.55 & 24:23:13.30 \\
2 & 13.71 & 0.51 & 3:43:09.76 & 24:41:32.82 \\
3 & 13.75 & 0.53 & 3:51:19.05 & 24:10:13.08 \\
... & ... & ...  & ...        & ...         \\
111 & 17.21 & 0.77 & 3:52:5.82 & 24:17:31.16 \\
112 & 17.34 & 0.84 & 3:48:50.45 & 25:17:54.52 \\
\hline
\end{tabular}
        \caption{Stellar candidates identified from our survey. The whole
          electronic list can be found on the CDS website.}
        \label{starcoo}
\end{table}

\subsection{Substellar domain}
\subsubsection{Photometric selection}

\begin{figure*}[htbp]
  {\centering
  \leavevmode
    \includegraphics[width=0.9\hsize]{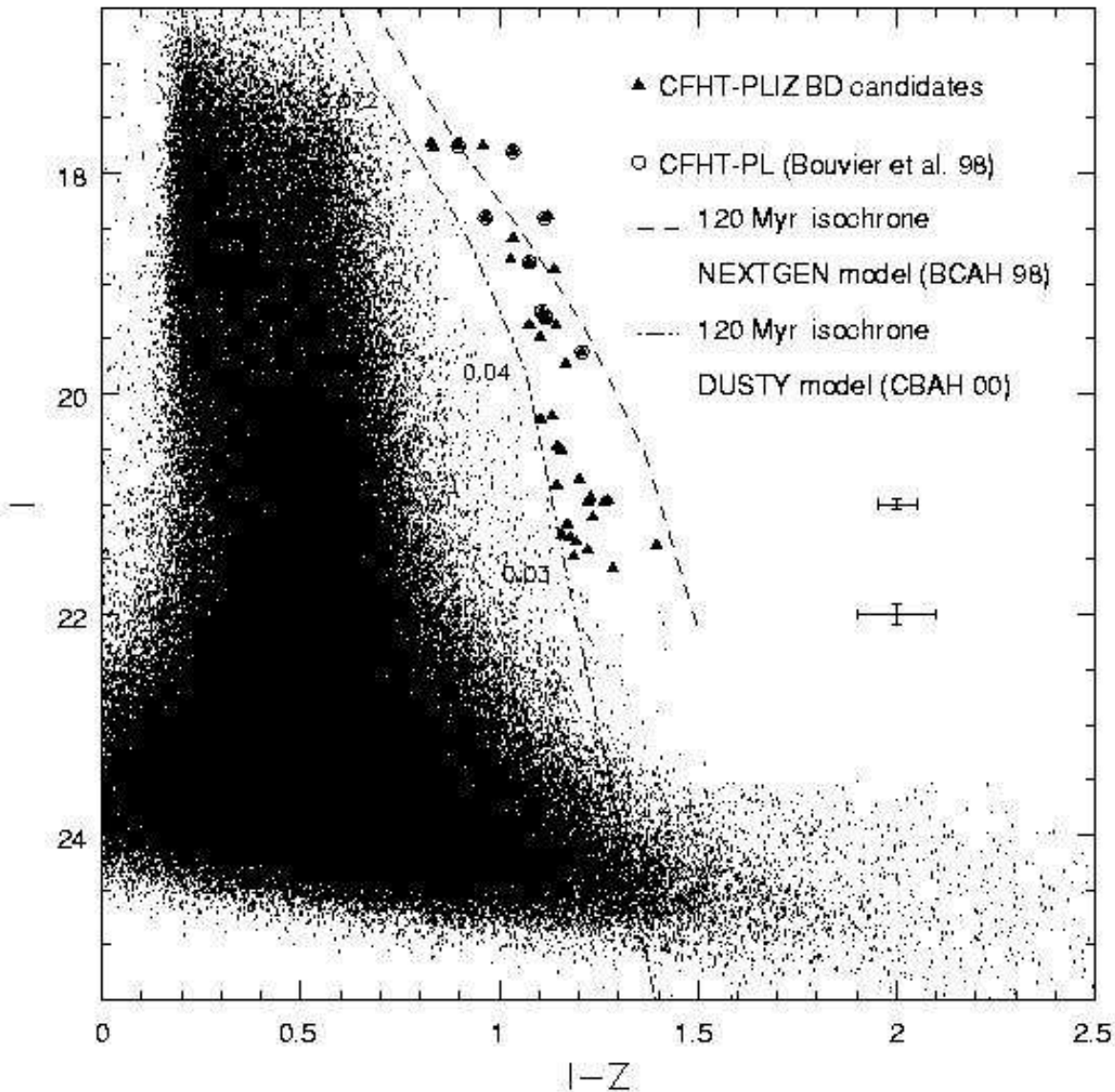}
        \caption{$(I,I-Z)$ color magnitude diagram for the long 
          exposures. The small dots represent the field stars. Brown
          dwarf candidates down to $0.03\msun$ are shown as filled
          triangles. Previously identified Pleiades proper motion BDs
          from Moraux et al. 2001 recovered by our survey are shown as
          open circles. The 120 Myr NEXTGEN (dashed line) and DUSTY
          (dot-dashed line) isochrones from Baraffe et al.  (1998) and
          Chabrier et al. (2000) are also shown. Error bars indicate
          the rms photometric error. }
        \label{cmdlong}}
\end{figure*}

The $(I,I-Z)$ color magnitude of the point-like objects contained in
the long exposures of the 17 Pleiades fields is shown in
Fig~\ref{cmdlong}.  Candidates previously identified by Bouvier et al.
(1998) and confirmed on the basis of spectroscopic data, infrared
photometry (Mart\'{\i}n et al. 2000) and proper motion (Moraux et al.
2001) are shown as open circles. These objects define the high mass
part of the cluster substellar sequence from $I\simeq17.8$ down to
about $I\simeq19.5$. We note that the location of two Pleiades members
(CFHT-PL-12 and CFHT-PL-16) suggest that they are likely binaries as
already suspected by Bouvier et al. (1998) and Mart\'{\i}n et al.
(2000).

We overplotted the 120 Myr isochrones from the NEXTGEN and DUSTY
models from Baraffe et al. (1998) and Chabrier et al. (2000)
respectively, assuming a distance modulus for the Pleiades cluster of
$(m-M)_{o}= 5.53$, $A_{V}=0.12$ and a solar metallicity.  At a
T$_{eff}$ which corresponds to late-M and early L spectral types, dust
grains begin to form, changing the opacity and resulting in objects
having bluer $I-Z$ colors than predicted by the NEXTGEN models.  DUSTY
models instead include a treatment of dust grains in cool atmospheres
for T$_{eff} \le$ 2300K. To build our sample of Pleiades brown dwarf
candidates for $17.8\le I\le 19$, we defined a line 0.1 mag bluer in
$I-Z$ than the NEXTGEN isochrone and we selected all the sources
located on the right side of this line. For $I\ge19$, we selected all
the objects redward of the DUSTY isochrone and we stopped our
selection around the completeness limit, i.e.  $M\simeq0.03\msun$. All
the candidates are shown in Fig~\ref{cmdlong} as filled triangles. The
photometry and coordinates of these objects are given in
Table~\ref{bdcoo}. Two sources are located $\sim$0.12 mag left on the
NEXTGEN isochrone at about $I=18.5$ and have not been considered as
brown dwarf candidates. The proper motion of the faintest of these two
objects has been measured and indicates non-membership.  The other
source will be followed up but this will not change the mass function
estimate.

\begin{table*}[htbp]
        \centering \leavevmode
\begin{tabular}{llllll}
\hline
%
CFHT-PLIZ & $I$ & $I-Z$ & $RA_{J2000}$ & $DEC_{J2000}$ & Other Id.\\ 
    &     &       & (h m s)    & (${}^{o}$ ' '') &  \\
\hline 
1 & 17.79 & 0.83 & 3:51:05.98 & 24:36:17.09 & \\ 
\bfseries 2 & 17.81 & 0.90 & 3:55:23.07 & 24:49:05.01 &  BPL 327 ($I_{KP}$=17.72), IPMBD 11 ($I_{C}$=18.07)\\ 
\bfseries 3 & 17.82 & 0.90 & 3:52:06.72 & 24:16:00.76 & CFHT-Pl-13 ($I_{C}$=18.02), Teide 2 ($I$=17.82), BPL 254 ($I_{KP}$=17.59)\\ 
4 & 17.82 & 0.96 & 3:41:40.92 & 25:54:23.00 & \\ 
5 & 17.84 & 0.84 & 3:53:37.96 & 26:02:19.67 & \\ 
\bfseries 6 & 17.87 & 1.04 & 3:53:55.10 & 23:23:36.41 &  CFHT-Pl-12 ($I_{C}$=18.00), BPL 294 ($I_{KP}$=17.61) \\ 
7 & 18.46 & 1.12 & 3:48:12.13 & 25:54:28.40 & \\ 
\bfseries 8 & 18.47 & 0.96 & 3:43:00.18 & 24:43:52.13 &  CFHT-Pl-17 ($I_{C}$=18.80), BPL 49 ($I_{KP}$=18.32) \\ 
\bfseries 9 & 18.47 & 1.11 & 3:44:35.19 & 25:13:42.34 & CFHT-Pl-16 ($I_{C}$=18.66) \\ %
\bfseries 10 & 18.66 & 1.03 & 3:51:44.97 & 23:26:39.47 & BPL 240 ($I_{KP}$=18.45) \\ 
\small (11) & 18.85 & 1.03 & 3:44:12.67 & 25:24:33.62 & CFHT-Pl-20 ($I_{C}$=18.96) \\ 
\bfseries 12 & 18.88 & 1.07 & 3:51:25.61 & 23:45:21.16 & CFHT-Pl-21 ($I_{C}$=19.00), Calar 3 ($I$=18.73), BPL 235 ($I_{KP}$=18.66)\\ 
13 & 18.94 & 1.14 & 3:55:04.40 & 26:15:49.32 & \\ 
14 & 18.94 & 1.14 & 3:53:32.39 & 26:07:01.20 & \\ 
\bfseries 15 & 19.32 & 1.11 & 3:52:18.64 & 24:04:28.41 & CFHT-Pl-23 ($I_{C}$=19.33)\\ 
\bfseries 16 & 19.38 & 1.12 & 3:43:40.29 & 24:30:11.34 & CFHT-Pl-24 ($I_{C}$=19.50), Roque 7 ($I$=19.29), BPL 62 ($I_{KP}$=19.19)\\ 
17 & 19.44 & 1.08 & 3:51:26.69 & 23:30:10.65 & \\ 
18 & 19.45 & 1.14 & 3:54:00.96 & 24:54:52.91 & \\ 
19 & 19.56 & 1.10 & 3:56:16.37 & 23:54:51.44 & \\ 
\bfseries 20 & 19.69 & 1.21 & 3:54:05.37 & 23:33:59.47 &  CFHT-Pl-25 ($I_{C}$=19.69), BPL 303 ($I_{KP}$=19.43)\\ 
21 & 19.80 & 1.17 & 3:55:27.66 & 25:49:40.72 & \\ 
22 & 20.27 & 1.13 & 3:51:52.71 & 26:52:32.16 & \\ 
\bfseries 23 & 20.30 & 1.10 & 3:51:33.48 & 24:10:14.16 & \\ 
24 & 20.55 & 1.15 & 3:47:23.68 & 26:00:59.75 & \\ 
\small (25) & 20.58 & 1.16 & 3:52:44.30 & 24:24:50.04 & \\ 
\bfseries 26 & 20.85 & 1.20 & 3:44:48.66 & 25:39:17.52 & \\ 
\small (27) & 20.90 & 1.14 & 3:55:00.38 & 23:38:08.05 & \\ 
28 & 21.01 & 1.23 & 3:54:14.03 & 23:17:51.39 & \\ 
29 & 21.03 & 1.27 & 3:49:45.29 & 26:50:49.88 & \\ 
30 & 21.04 & 1.22 & 3:51:46.00 & 26:49:37.41 & \\ 
31 & 21.05 & 1.26 & 3:51:47.65 & 24:39:59.51 & \\ 
32 & 21.19 & 1.23 & 3:50:15.47 & 26:34:51.27 & \\ 
33 & 21.25 & 1.17 & 3:50:44.68 & 26:42:09.36 & \\ 
34 & 21.35 & 1.16 & 3:54:02.56 & 24:40:26.07 & \\ 
35 & 21.37 & 1.18 & 3:52:39.17 & 24:46:30.03 & \\ 
36 & 21.42 & 1.19 & 3:54:38.34 & 23:38:00.63 & \\ 
37 & 21.45 & 1.40 & 3:55:39.57 & 24:12:52.12 & \\ 
38 & 21.49 & 1.22 & 3:45:54.69 & 26:30:14.57 & \\ 
39 & 21.55 & 1.19 & 3:53:40.30 & 26:16:18.15 & \\ 
40 & 21.66 & 1.29 & 3:49:49.30 & 26:33:56.19 & \\ 
\hline
\end{tabular}
        \caption{Brown dwarfs candidates identified from our survey. 
          Objects written in bold characters (resp. in parenthesis) 
          have proper motion indicating cluster membership (resp. non 
          cluster membership).}
        \label{bdcoo}
\end{table*}

Part of our survey overlaps with Bouvier et al.'s (1998) survey
performed in 1996, so that we were able to derive proper motion for
some of the objects identified in both surveys. The two epochs of
observations are separated by approximately 4 years and the resulting
proper motion uncertainty is typically $1\sigma\simeq 7$mas/yr. The
details of the procedures used to derive proper motion are given in
Moraux et al. (2001). Objects which have a proper motion less than
$2\sigma$ from the cluster motion ($\mu_{\alpha}=+19$ mas/yr,
$\mu_{\delta}=-43$ mas/yr) are very likely Pleiades members. We have
written those objects in bold characters in Table~\ref{bdcoo} and
objects whose proper motion indicates non membership in parenthesis.\\
Two of our new candidates had already been identified as probable
Pleiades members by Pinfield et al. (2000), CFHT-PLIZ-2=BPL 327 and
CFHT-PLIZ-10=BPL 240, on the basis of their optical and infrared
photometry and from their proper motion by Hambly et al. (1999).
These objects are also written in bold characters in
Table~\ref{bdcoo}.

\subsubsection{Contamination}

An ($I,I-Z$) diagram alone cannot identify objects as certain Pleiades
members as one expects some level of contamination by field stars. Due
to the relatively high galactic latitude of the Pleiades cluster,
heavily reddened distant objects should not contaminate our
photometric sample of candidate members. However, the relative
location in the CMD of the theoretical Pleiades isochrone and a zero
age main sequence isochrone from DUSTY models indicates that some of
the photometrically selected brown dwarfs candidates could in fact be
field M-dwarfs at a distance about 30\% closer than the Pleiades.
Considering the whole selection range which extends from 0.5 mag below
the cluster sequence to the binary cluster sequence, we find that
contaminating field M-dwarfs can lie in a distance range from 60 to
125 pc. Then, taking into account the area of the survey, the volume
occupied by contaminants is about 1150 pc$^{3}$.  The field star
luminosity function for $M_{I}=12-14.5$ can be approximated as a
constant $\phi \sim 0.003$ stars/pc$^{3}$ per unit $M_{I}$ as
estimated from the DENIS survey (Delfosse 1997).  We therefore expect
to find about 7 field stars out of 21 candidates in the range
$I=17.8-19.8$, i.e. a contamination level of about 33\% as previously
derived from proper motion measurements by Moraux et al. (2001).

At fainter magnitudes the contamination level cannot be derived from
the field star luminosity function which is not very well known for
$M_{I}>14.5$. However, we can use the number of stars identified in
the DENIS survey down to $I=18$ in a given color (or temperature)
range in order to estimate the number of contaminants in our sample
for this color interval. For example, the brown dwarf candidates with
$I$ between 20.2 and 21.7 have a temperature of $\sim2000$K (Chabrier
et al. 2000). We then consider the number of DENIS objects within a
restricted temperature range around this value and $I$ between 16.5
and 18, and multiply this number by two factors~: a) the ratio of our
CFHT survey area to the DENIS survey area, and b) the ratio of the two
volumes corresponding to the two magnitude ranges ($I=$20.2 to 21.7
and $I=$16.5 to 18) for the DENIS survey. We thus predict 5 or 6 field
dwarfs to occupy the region of the Pleiades color magnitude diagram
corresponding to $0.04\msun \ge M\ge 0.03\msun$.  This indicates again
a contamination level of $\sim 30\%$.

For redder objects, the statistics of the DENIS survey is very low so
that it is difficult to estimate the contamination. Moreover, our
survey starts to be incomplete in this domain and that is why we
limited our analysis to $M=0.03\msun$.

\subsection{Radial distribution}

In order to deduce the total number of Pleiades members and derive the
cluster mass function from a survey covering only a fraction of the
cluster area, we need to investigate the spatial distribution of
cluster members and its dependence on mass.

\begin{figure*}[htbp]
        \centering \leavevmode
        \parbox{0.5\hsize}{
          \includegraphics[width=\hsize]{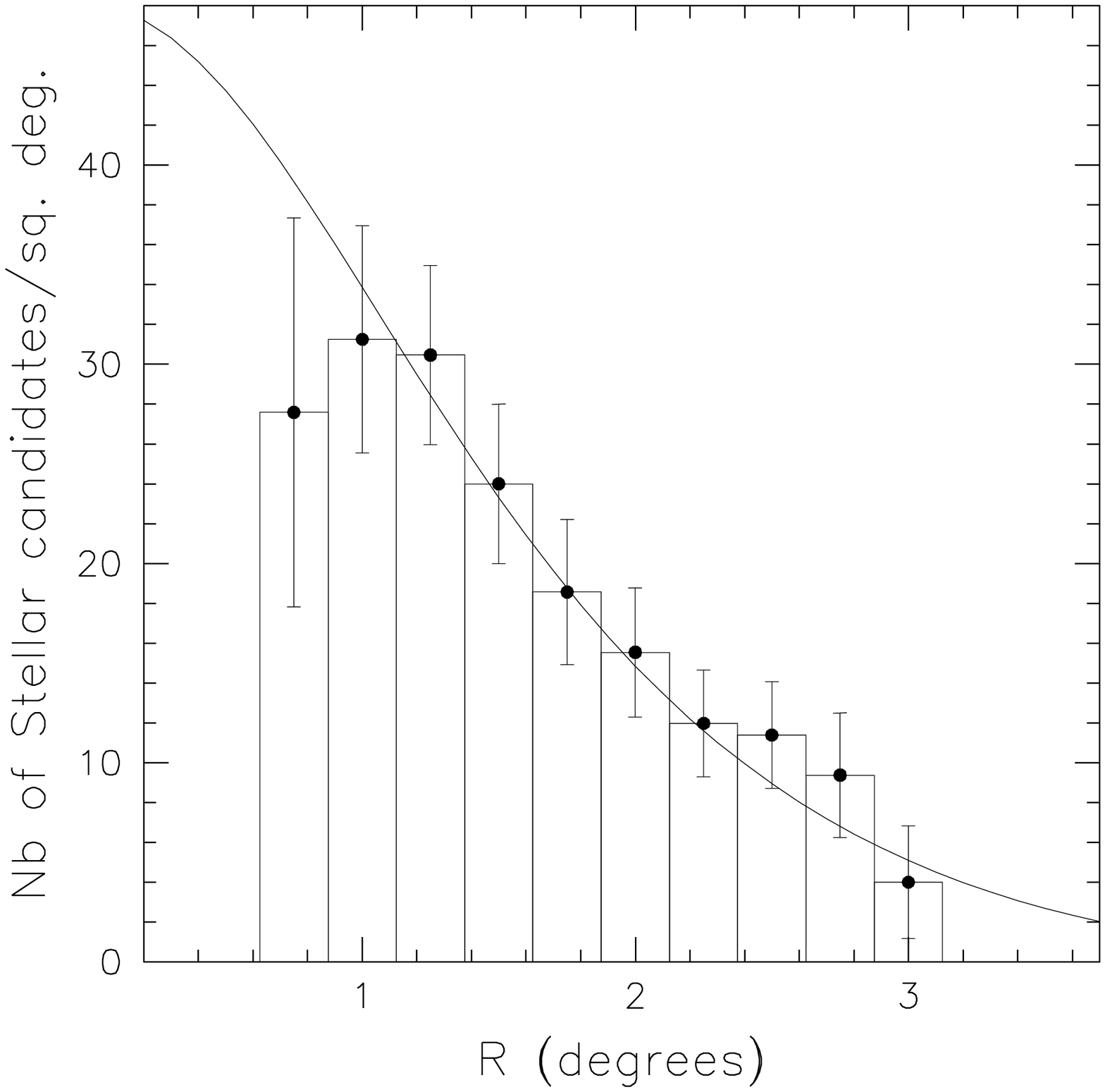}
          }\hfill
        \parbox{0.5\hsize}{
          \includegraphics[width=\hsize]{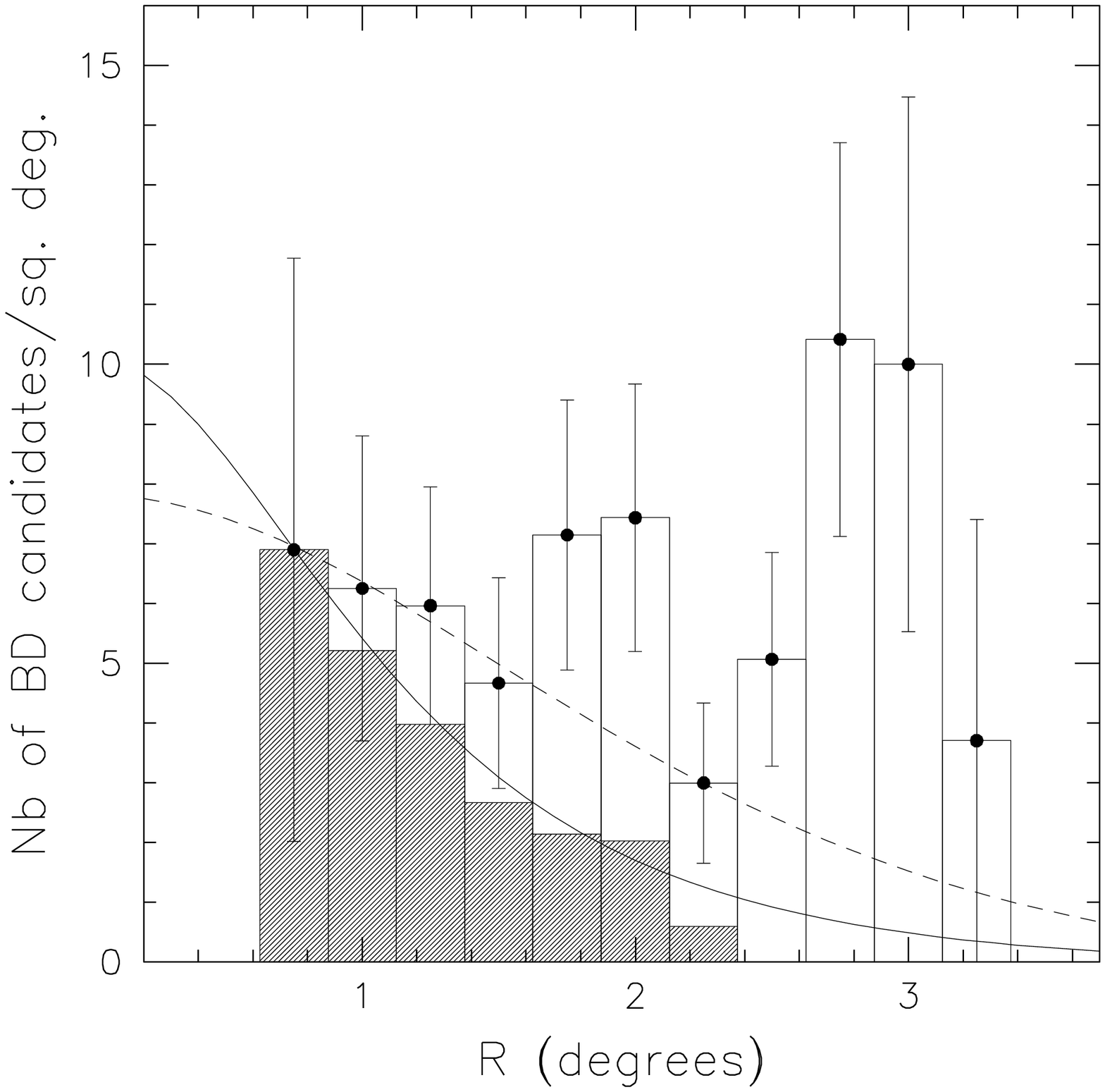}
          }
        \caption{$Left~:$ The radial distribution of probable Pleiades
          stellar members found in our survey and having a mass
          between $0.48\msun$ and $0.08\msun$ (histogram). Overplotted
          is the best King profile fit which, assuming
          $r_{t}=5.54^{o}$ (Pinfield et al. 1998), yields a core
          radius $r_{c}=2.0^{o}$; $Right~:$ The plain histogram shows
          the radial distribution of our brown dwarf candidates and
          the shaded histogram represents those which are already
          confirmed by proper motion.  Overplotted are King profiles
          with $r_{t}=5.54^{o}$.  The solid line is a fit of the
          shaded histogram which yields a lower limit of
          $r_{c}=1.3^{o}$ for the substellar core radius. A King
          profile with a core radius of $r_{c}=3.0^{o}$ is shown for
          reference (dashed line). }
        \label{distrib}
\end{figure*}

The spatial distribution of our Pleiades brown dwarf candidates is
shown in Fig.~\ref{fieldlocation}. Overplotted are circles of radii
0.75 to 3.5 degrees centered on the cluster center.  From this diagram
we estimated the covered area within annulii of $0.25\degr$ width and
we counted the number of substellar objects found therein. We then
obtained radial surface densities for brown dwarf candidates by
dividing these numbers by the corresponding surveyed areas. We
proceeded in the same way for low mass stars. The number of stellar
and substellar objects per square degree as a function of the radial
distance is shown in Fig~\ref{distrib}.

The stellar distribution ($13.5<I<17.5$, i.e. $0.48>M>0.08\msun$)
shown on the left panel is well fitted by a King distribution (King
1962)~:
  \begin{equation}
    \label{eq:1}
    f(x) = k\,\left[\frac{1}{\sqrt{1+x}} - 
      \frac{1}{\sqrt{1+x_{t}}}\right]^2
  \end{equation}
where $k$ is a normalisation constant, $x=(r/r_{c})^2$ and
$x_{t}=(r_{t}/r_{c})^2$ with r the radius from the cluster center.
The core radius $r_{c}$ increases as the stellar mass decreases and
the tidal radius $r_{t}$ corresponds to the location where the
gravitationnal potential of the galaxy equals the cluster potential.
Using $r_{t}=5.54\degr$ from Pinfield et al. (1998), we found
$k\simeq110$ per square degrees and $r_{c}\simeq2$ degrees for a
median mass of our stellar sample of $M\simeq0.2\msun$. From this
radial distribution, we obtain a total of 557 stars between 0.08 and
0.48$\msun$.  For a dynamically relaxed cluster, the core radius is
expected to vary with stellar mass as $M^{-0.5}$ and Jameson et al.
(2002) derived the relationship $r_{c}=0.733M^{-0.5}$ for the
Pleiades stars. For $M\simeq0.2\msun$, this yields
$r_{c}\simeq1.6\degr$, slightly smaller than our value.

The radial distribution of brown dwarf candidates is shown on the
right side of the figure~\ref{distrib}. The plain histogram
corresponds to the whole list of candidates whereas the shaded
histogram corresponds to objects having proper motion consistent with
cluster membership (written in bold characters in Table~\ref{bdcoo}).
This histogram does not extend further than $r=2.25^{o}$ corresponding
to the surveys of Pinfield et al. (2000) and Bouvier et al. (1998)
from which proper motions have been derived.  Those surveys were not
as deep as ours so that only brown dwarf candidates brighter than
$I=20.9$ were counted.  A King profile fitted to this histogram yields
$r_{c}\simeq1.3$ degrees as a lower limit to the cluster substellar
core radius.  The plain histogram also decreases in the first few
radius bins but then increases further away from the cluster center.
Note, however, that the uncertainties due to small number statistics
are large and the plain histogram is not corrected for contamination
for field stars, whose rate is expected to increase away from the
cluster center. An illustrative King profile with $r_{c}\simeq3.0$
degrees is shown as a possible fit to the brown dwarf distribution,
mainly based on the few first radial bins. The median mass for the
brown dwarfs candidates is $M\simeq0.05\msun$ and the value expected
by Jameson et al.'s (2002) relationship $r_{c} = 0.733 M^{-0.5}$ is
3.4 degrees.

With $r_{c}\simeq3.0^{o}$ and $k=28.5$ per square degrees, integration
of the King distribution yields a total of $\sim130$ brown dwarfs
between $17.8<I<21.7$, i.e. between 0.07 and 0.03$\msun$ in the whole
cluster.  From this distribution, we expect to find $\sim26$ brown
dwarfs Pleiades members in our survey out of the 40 selected
candidates. This would correspond to a contamination level of 35\%,
quite consistent with our estimate above.

\subsection{The Pleiades mass function}


The Pleiades mass function can be estimated from our CFHT large survey
over a continuous mass range from $0.03\msun$ to $0.45\msun$.  For the
stellar part (down to $I=17.5$) we use our sample of candidates
derived from short exposure images and decontaminated as explained
above. We derived masses from $I$-band magnitudes using the 120 Myr
isochrone from Baraffe et al. (1998).  Below the HBML ($I\sim17.5$) we
consider our selection of brown dwarf candidates (Table~\ref{bdcoo})
and we apply a correction factor of 0.7, assuming a contamination
level of 30\%. We used the 120 Myr isochrone from the DUSTY models of
Chabrier et al. (2000) to estimate masses.

In order to correctly estimate the mass function of the whole cluster
from a survey which is spatially uncomplete, one has to take into
account the different radial distribution of low mass stars and brown
dwarfs. Our CFHT fields are located between 0.75 and 3.5 degrees from
the cluster center. We now proceed to estimate the fraction of low
mass objects located in this ring compared to the total number of such
objects in the cluster. For a King-profile surface density
distribution, the total number of stars seen in projection within a
distance $r$ of the cluster center is obtained by integrating
equation~\ref{eq:1}~:
  \begin{equation}
    \label{eq:2}
    n(x)= k\pi r_{c}^{2} \left[\ln(1+x) - 4\frac{\sqrt{1+x}-1}{\sqrt{1+x_{t}}}
      + \frac{x}{1+x_{t}}\right]
  \end{equation}
We consider stellar core radii following Jameson et al. (2002)
relationship $r_{c} = 0.733 M^{-0.5}$ and assumed a substellar core
radius $r_{c}=3.0^{o}$ (see previous section). We then deduce from
equation~\ref{eq:2} that 74\% of the $0.4\msun$ stars and 80\% of
the brown dwarfs are located within a distance from the cluster
center between 0.75 and 3.5 degrees. This indicates that the
relative number of brown dwarfs compared to low mass stars deduced
from our survey is representative to their relative number over the
whole cluster. In other words, the area covered by this survey is
large enough so that any correction to the mass function for a
mass-dependent radial distribution is negligeable.

\begin{figure}[htbp]
        \centering \leavevmode
        \includegraphics[width=\hsize]{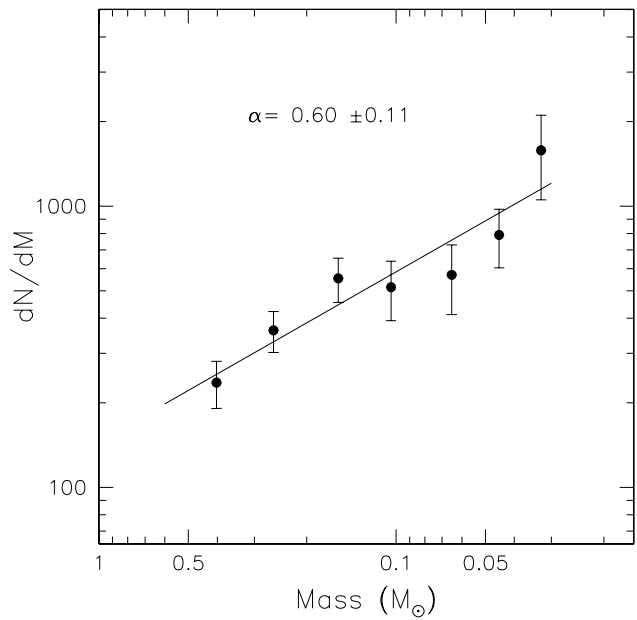}
        \caption{The Pleiades mass function between $0.03\msun$ and
          $0.45\msun$.  The dots corresponds to the number of objects
          per unit mass found in our survey. The last three points
          corresponds to the substellar domain and have been corrected
          for a 30\% contamination level by field stars (see text).
          The data points are fitted by a power law with an index
          $\alpha=0.60\pm 0.11$ ($dN/dM \propto M^{-\alpha}$).}
        \label{imf}
\end{figure}

The derived mass function is shown in Fig.~\ref{imf} as the number of
objects per unit mass.  Within the uncertainties, it is reasonably
well-fitted by a single power-law $dN/dM \propto M^{-\alpha}$ over the
mass range from $0.03\msun$ to $0.45\msun$. A possibility to explain
why the $0.06\msun$ data point is low is discussed in Dobbie et al.
(2002). A linear regression through the data points yields an index of
$\alpha=0.60\pm0.11$, where the uncertainty is the $1\sigma$ fit
error\footnote{Using the lower limit of $r_{c}=1.3\degr$ for the
  cluster core radius in the susbtellar domain would yield
  $\alpha=0.63\pm0.11$ instead.}. This result is consistent with
previous estimates (cf.  Table~\ref{bdsurveys}) and is affected by
smaller uncertainties thanks to the combination of relatively large
samples of low mass stars and brown dwarfs, proper correction for
contamination by fields stars, and extended radial coverage of the
cluster.

In Moraux et al. (2001) the mass function index was found to be
$\alpha=0.51\pm0.15$. Lacking a proper determination of the radial
distribution of cluster members, the assumption was made that brown
dwarfs and very low mass stars were similarly distributed. The index
estimate was based on Bouvier et al.'s (1998) survey which covered
fields spread between 0.75 and 1.75 degrees from the cluster center.
From the radial distribution derived above, we find that this annulus
contains 43\% of the $0.4\msun$ stars and 33\% of the cluster brown
dwarfs. Applying these correcting factors to the number of Pleiades
members found in that survey, the mass function index becomes 0.63
instead of 0.51. This corrected value is in excellent agreement with
our new estimate.

Extrapolating the power-law mass function down to $M=0.01\msun$, we
predict a total number of $\sim270$ brown dwarfs in the Pleiades for a
total mass of about $10\msun$. Clearly, while brown dwarfs are
relatively numerous, they do not contribute significantly to the
cluster mass.  Adams et al. (2001) derived a total mass of
$\sim800\msun$ for the Pleiades which means that, even though brown
dwarfs account for about 25\% of the cluster members, they represent
less than 1.5\% of the cluster mass.

\section{Discussion}

One of the main motivations for the determination of the lower mass
function is to constrain the star and brown dwarf formation processes.
A pressing issue is then whether the Pleiades mass function (MF)
observed at an age of $\sim120$ Myr is representative of the initial
mass function (IMF), i.e. the mass spectrum resulting from the
formation process.

We compare our results to the mass function of other young open
clusters and star forming regions in order to investigate the
dynamical evolution of the Pleiades cluster from a few Myr to its
present age of 120 Myr. We also compare the Pleiades MF to the
Galactic disk mass function in order to constrain the brown dwarf
formation process.

\begin{figure}[htbp]
        \centering
        \leavevmode
          \includegraphics[width=\hsize]{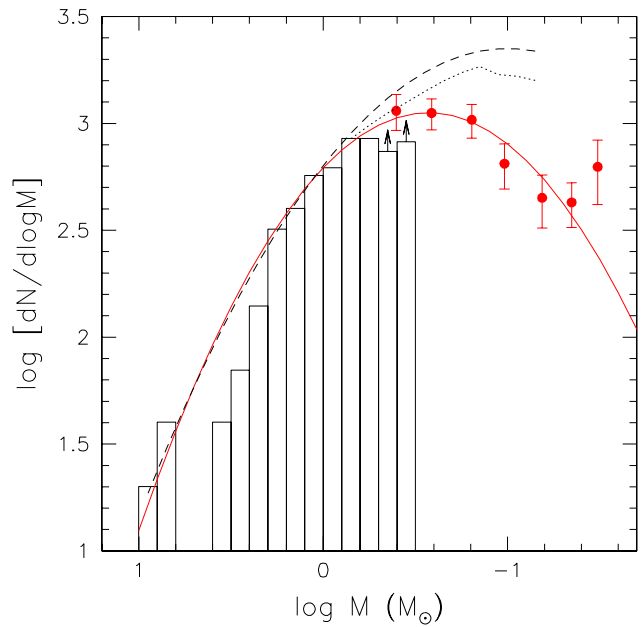}
        \caption{The Pleiades mass function represented as the number
          of objects per logarithmic mass units over the mass range
          $0.03\msun\le M\le 10\msun$. In this representation
          Salpeter's slope is 1.35.  The histogram corresponds to the
          Pleiades star catalog built from the Prosser and Stauffer
          Open Cluster Database$^3$ and the large dots are our data
          points.  The Pleiades mass function is fitted by a
          log-normal function (solid line) over the entire mass range.
          The dashed line corresponds to the Galactic disk mass
          function from Chabrier (2001) and the dotted line to the
          estimated Pleiades mass function corrected for unresolved
          binaries.}
        \label{mf_tot}
\end{figure}

\subsection{Comparison to other young clusters~: clues to early dynamical 
evolution}

We compare our results to those obtained recently for other young open
clusters such as M35 ($\sim150$ Myr, Barrado et al. in prep) and
$\alpha$~Per ($\sim80$ Myr, Barrado et al. 2002). The lower mass
function of those clusters can be approximated by a single power-law
with an index $\alpha=0.58$ between $0.4$ and $0.1\msun$ for M35 and
$\alpha=0.56$ between $0.2$ and $0.06\msun$ for $\alpha$~Per. These
values are very similar to the one determined here for the Pleiades
($\alpha=0.60\pm0.11$).  Results obtained for star forming regions
such as $\sigma$-Orionis ($\sim5$ Myr, Bejar et al. 2001) and IC348
($\sim3$ Myr, Tej et al. 2002) are also consistent with this value.
The power-law indices are $\alpha=0.8\pm0.4$ ($0.2>M>0.013\msun$) for
$\sigma$-Orionis and $\alpha=0.7\pm0.2$ ($0.5>M>0.035\msun$) for
IC348. The mass functions of Pleiades-age clusters and star forming
regions thus appear to have a similar shape across the
stellar-substellar boundary.

Comparison over a larger mass domain is best achieved by plotting the
number of stars per logarithmic mass units~:
  \begin{equation*}
    \label{eq:3}
    \xi(\log M) = \frac{dN}{d\log M}
  \end{equation*}
(in this representation, $dN/dM\propto M^{-1.0}$ corresponds to
$dN/d\log M$ constant). Fig.~\ref{mf_tot} shows the Pleiades mass
function over the mass range from 0.03$\msun$ to 10$\msun$.  The
stellar portion of the mass function has been derived from the
Prosser and Stauffer Open Cluster Database\footnote{Available at
   http://cfa-www.harvard.edu/stauffer/opencl/} constructed from
several proper motion surveys. This catalog, however, becomes
uncomplete below $\sim 0.5\msun$. We therefore normalized our survey
so that the total number of objects over the 0.08-0.48$\msun$ mass
range corresponds to the number computed from the King profile (see
section 3.3). This normalization ensures the continuity of the
Pleiades mass function shown in Fig.~\ref{mf_tot} from the brown
dwarfs up to the most massive stars of the cluster.

The cluster mass function is fitted by a lognormal form over more than
2 decades in mass~:
\begin{equation}
  \label{eq:4}
  \xi(\log M) \propto \exp\left[ - \frac{(\log M - \log \langle M\rangle)^2}
    {2\sigma_{logM}^{2}}\right]
\end{equation}
with $\langle M\rangle\simeq0.25\msun$ and $\sigma_{logM}=0.52$.
This mass function can be approximated by a single power-law above
$1.5\msun$, $\xi(\log M)\propto M^{-1.7}$ in logarithmic mass units (or
$dN/dM\propto M^{-2.7}$ in linear mass units), and peaks at $M\simeq
0.25\msun$ before decreasing at lower mass.  Luhman et al. (2000)
found that $\rho$~Ophiuchus, IC348 and the Trapezium mass functions
$\xi(\log M)$ also exhibit a maximum around $0.25\msun$ and are quite
similar to the Pleiades one in the stellar domain.

Within uncertainties, we thus find no evidence for significant
differences in the mass function of the Pleiades, other Pleiades-age
open clusters and star forming regions across the stellar-substellar
boundary but also over the entire stellar domain.  This suggests that
the cluster population observed at an age of about 120 Myr is still
quite similar to its population at only a few Myr. As a cluster
evolves weak gravitational encounters occur leading eventually to the
evaporation of the lowest mass members. The results above seem to
indicate that such a dynamical evolution has not yet affected the mass
function of the cluster.  This is in agreement with dynamical models
(e.g. de la Fuente Marcos \& de la Fuente Marcos 2000, Adams et al.
2002) which predict that only about 10\% of the total number of
cluster members is lost after $\sim100$ Myr, and that the fraction of
brown dwarfs to stars remains nearly constant, so that the shape of
the mass function across the stellar-substellar boundary is hardly
affected.

The mass function of the Pleiades at an age of $\sim120$ Myr thus
seems to be representative of the cluster population at an early age
of a few Myr. But does it also reflects the initial population of the
cluster at the time it formed (IMF)?  Rapid dynamical processes
associated to cluster formation might conceivably lead to the prompt
ejection of low mass stars and brown dwarfs during the very early
stage of a cluster life, before a few Myr. If instrumental, such
processes might result in a depletion of the low mass cluster
population very early on and thus modify the shape of the mass
function.

An example of conditions under which such processes may occur is when
the gas is expelled from the cluster during its formation. The
gravitational potential decreases drastically and all the low mass
objects in the outer part of the cluster are ejected. However, current
models (Kroupa 2001) predict that roughly as many massive stars as low
mass stars are ejected in this process, unless mass segregation is
present initially. Hence, the shape of the mass function should not be
significantly affected.

Another violent dynamical process may be associated to the formation
of brown dwarfs themselves.  Reipurth \& Clarke (2001) proposed that
substellar objects are ``stellar embryos'' according to the following
scenario~: as molecular cloud cores fragment to form unstable
protostellar multiple systems which decay dynamically, the lowest mass
fragments are ejected from the core, and deprived of surrounding
gas to accrete they remain substellar objects. In this scenario, the
velocity dispersion may be expected to be larger for brown dwarfs than
for stars and, for a fraction of substellar objects, the velocity may
indeed exceed the escape velocity. Such a process could then be very
efficient in quickly removing the lowest mass objects from the
cluster, thus strongly modifying its initial mass function. A
possibility to constrain this brown dwarf formation process would be
to search for an observational signature of their primordial
kinematics. Unfortunately, at an age of 120~Myr the Pleiades cluster
is already largely relaxed and has lost memory of its initial
evolution.

Furthermore, while several models have been developped to investigate
the dynamical ejection of brown dwarfs, predictions regarding their
initial velocity distribution differ. Sterzik \& Durisen (1998) find
that the velocity dispersion depends on both mass and binarity, being
larger for low mass single objects than for massive binaries.
Delgado-Donate et al. (2002) find that it depends only on binarity and
that nearly all ejected objects are preferentially low mass and single
ones. According to M.~Bate (priv. comm.) the velocity
dispersion depends neither on mass nor on binarity so that dynamical
ejection of low mass objects should not affect the overall mass
function of the cluster. Pending the resolution of these
uncertainties, it is then difficult to assess whether the proposed
ejection mechanism would lead to peculiar kinematical signatures in
the substellar population and would thus possibly affect the cluster
mass function. This issue is further discussed in the next section by
comparing the Pleiades and the Galactic disk mass functions.

\subsection{The Pleiades and Galactic disk mass functions~: clues to the brown dwarf formation process}

Below $\sim0.8\msun$ field stars have not had time to evolve off the
main sequence, so that the observed Galactic disk mass function ought
to be representative of the initial mass function. A log-normal fit to
the Galactic disk mass function in the mass range 0.1-1.0$\msun$ has
been derived by Chabrier (2001) based on large scale photometric
surveys of low mass stars in the solar neighbourhood. This log-normal
approximation to the field mass function is shown as a dashed line in
Fig.~\ref{mf_tot} where it has been normalized so as to have the same
number of $1\msun$ stars as in the Pleiades. Major differences clearly
appear between the Pleiades and the field mass functions at low
masses. The log normal mass function peaks at $M\simeq 0.1\msun$ for
the disk population and at $\simeq 0.25\msun$ for Pleiades members.
Furthermore, Chabrier (2002) estimates that the brown dwarf
population in the disk is comparable in number to the stellar one,
$N_{BD} \simeq N_{*}$, and that the substellar mass contribution to
the disk budget amounts to about $\sim 10\%$. In the Pleiades, we have
instead $N_{BD} \simeq N_{*}/3$ and a brown dwarf mass contribution to
the cluster mass of only $1.5\%$.  Indeed, the Pleiades mass function
lies well below the disk's one at the stellar-substellar boundary in
Fig.~\ref{mf_tot}.

Part of the observed difference may arise from the effect of binarity,
which is not accounted for in the Pleiades mass function. While the
Pleiades mass function derived above includes unresolved cluster
binaries, the Galactic disk mass function has been derived for the
single star population (i.e. all binaries are resolved, cf. Chabrier
2001). In order to correct the observed Pleiades mass function for
unresolved binaries at low masses, we follow Luhman et al. (1998) who
used a Monte Carlo technique to estimate the difference between single
star and system mass functions. Assuming that the properties of field
binaries derived by Duquennoy \& Mayor (1991) apply to Pleiades
systems (cf. Bouvier, Rigaut \& Nadeau 1997), we form unresolved
binaries by randomly pairing objects drawn from a segmented power law
mass function which represents the single star mass distribution down
to the stellar-substellar boundary\footnote{We do not attempt to
  correct the observed Pleiades mass function for unresolved binaries
  in the substellar domain because the brown dwarf binary statistics
  is still very poorly known, as is the statistics of brown dwarfs
  companions to low mass stars}. For an assumed 50\% binary fraction
in the Pleiades cluster, we find no major differences between the star
and system mass functions at $M>0.6\msun$ since above this mass the
presence of a low mass companion does not significantly affect the
determination of the primary's mass.  Below $0.6\msun$ down to
$0.07\msun$, however, we find $\Delta\alpha\sim0.5$ between the power
law exponents of the single star and system mass functions.

Then, approximating the log-normal form of the Pleiades mass function
derived above by a three-segment power-law between $0.07\msun$ and
$0.6\msun$, we estimate the binary corrected Pleiades mass function
which is shown in Fig.~\ref{mf_tot} (dotted line).  The binary
corrected Pleiades mass function and the Galactic disk mass function
now both peak around $0.13-0.1\msun$ and their shapes are roughly
identical in the stellar domain down to $0.07\msun$.  This suggests
that the difference between the observed Pleiades and Galactic disk
mass functions in the stellar range is merely the result of unresolved
Pleiades binaries. When binarity is properly accounted for, the
stellar mass function of the Pleiades is consistent with that of field
stars of the solar neighbourhood. The comparison of the two mass
functions in the substellar domain cannot be as detailed, due to the
large uncertainties still affecting the derivation of the substellar
mass function in the Galactic disk. Chabrier (2002) derives an upper
limit of $\alpha\leq 1$ for a power law approximation of the field
substellar mass function (see also Reid et al. 1999). This upper limit
is consistent with the Pleiades substellar mass function exponent
$\alpha\simeq 0.6$ derived above.

The similarity of the Pleiades and field mass functions down to at
least the substellar limit and possibly below suggests that low mass
objects have remained in the cluster at the time of its formation.
Hence, we do not find evidence for massive ejection of the lowest mass
objects early in the life of the cluster, which suggests either that
dynamical ejection from protostellar groups is not the dominant mode
of brown dwarf formation, or that this process yields a velocity
dispersion for brown dwarfs which is not different from that of stars.

\section{Conclusion}

We have conducted a deep wide field photometric survey of the Pleiades
cluster to build a sample of probable cluster members with masses in
the range $0.03\msun$ to $0.48\msun$.  We have identified 40 brown
dwarfs candidates, of which 29 are new discoveries. Taking into
account the radial distribution of cluster members, we derive the
cluster mass function accross the stellar-substellar boundary. We find
that a single power-law $dN/dM \propto M^{-\alpha}$ with an index
$\alpha=0.60\pm0.11$ provides a good match to the cluster mass
function in the $0.03-0.48\msun$ range. This new estimate is based on
a survey which combines a large radial coverage of the cluster and a
realistic assessment of the contamination by field stars. Furthermore,
the survey completely covers the $0.03-0.48\msun$ mass range, so that
the result does not rely on the combination of heterogeneous surveys,
as has been the case before. We therefore believe this new estimate is
reasonably robust. Small changes may be expected when our survey has
been followed up with either infrared photometry and/or proper
motions.

Over a larger mass domain, covering almost 3 decades in masses from
$0.03\msun$ to $10\msun$, we find that the cluster mass function is
better fitted by a log-normal distribution with $\langle
M\rangle\simeq0.25\msun$ and $\sigma_{logM}\simeq 0.52$. When
unresolved Pleiades binaries are taken into account, the log-normal
Pleiades mass function is not unlike the Galactic disk mass function.
This suggests that the dynamical evolution of the cluster has had yet
little effect on its mass content at an age of 120~Myr. It also
suggests that the brown dwarf formation process does not lead to the
dynamical evaporation of substellar objects at the time the cluster
forms.

\begin{acknowledgements}
  We thank E.~Bertin for allowing us access to PSFex before public
  release, E.~Magnier for his help in the astrometric calibration of
  the frames, J.~Adams and N.~Hambly for providing us data in
  electronic form prior to publication, I.~Baraffe for computing
  specific substellar isochrones for us and K.~Luhman for providing
  his Monte-Carlo software program to estimate the effect of
  unresolved binaries on the mass function. We also gratefully
  aknowledge helpful discussions with X.Delfosse on estimating field
  star contamination from DENIS data, with M.~Bate, C.~Clarke,
  E.~Delgado and M.~Sterzik on the dynamical evolution of young brown
  dwarfs in clusters, and with G.~Chabrier who brought our attention
  on the Galactic disk mass function and on the effect unresolved
  binary systems might have on the shape of the observed mass
  function.

\end{acknowledgements}

\end{document}